\documentclass[dvipdfmx,aip,jmp,reprint,onecolumn]{revtex4-1}

\usepackage{amsmath,amsthm,amssymb,mathptmx,color}
\usepackage{bm,graphicx}
\usepackage{fancybox}

\newtheorem{definition}{Definition}[section]

\newtheorem{prop}[definition]{Proposition}

\begin{document}

\title{Antisymmetric tensor generalizations of affine vector fields}
\author{Tsuyoshi Houri}
\email[E-mail: ]{houri@phys.sci.kobe-u.ac.jp}
\affiliation{Department of Physics, Kobe University, 1-1 Rokkodai, Nada, Kobe, Hyogo, 657-8501 JAPAN}
\author{Yoshiyuki Morisawa}
\email[E-mail: ]{morisawa@keiho-u.ac.jp}
\affiliation{Osaka University of Economics and Law, 6-10 Gakuonji, Yao, Osaka, 581-8511 JAPAN}
\author{Kentaro Tomoda}
\email[E-mail: ]{k-tomoda@stu.kobe-u.ac.jp}
\affiliation{Department of Physics, Kobe University, 1-1 Rokkodai, Nada, Kobe, Hyogo, 657-8501 JAPAN}

\date{\today}

\begin{abstract}
Tensor generalizations of affine vector fields called symmetric and antisymmetric affine tensor fields
are discussed as symmetry of spacetimes. We review the properties of the symmetric ones,
which have been studied in earlier works, and investigate the properties of the antisymmetric ones,
which are the main theme in this paper. It is shown that antisymmetric affine tensor fields
are closely related to one-lower-rank antisymmetric tensor fields which are parallelly transported along geodesics.
It is also shown that the number of linear independent rank-$p$ antisymmetric affine tensor fields
in $n$ dimensions is bounded by $(n+1)!/p!(n-p)!$.
We also derive the integrability conditions for antisymmetric affine tensor fields.
Using the integrability conditions, we discuss the existence of antisymmetric affine tensor fields
on various spacetimes.
\end{abstract}

\pacs{02.40.-k,02.40.Ky,04.20.-q}

\keywords{Spacetime symmetry, affine collineation, affine vector field, parallel transport, pp-wave spacetime}

\preprint{KOBE-TH-15-10}

\maketitle

\section{Introduction}\label{sec:introduction}
Spacetime symmetry, i.e. isometry described by Killing vector fields (KVs),
has helped us in understanding of the nature of spacetimes.
It is widely known that KVs form Lie algebra of finite dimensions
with respect to the Lie bracket.
Hence, spacetimes have been classified by Lie algebras of KVs.
Afterwards, homothetic vector fields (HVs), conformal Killing vector fields (CKVs) and
affine vector fields (AVs), which also form Lie algebras of finite dimensions, 
have been discussed as spacetime symmetry.
Such symmetry vector fields have also played an important role in physics,
especially in general relativity,
in understanding of behaviors of matters in gravitational fields.
For example, if a spacetime possesses a (C)KV,
one can construct a conserved quantity along (null) geodesics.
On a spacetime having an AV, one obtains a Jacobi field, which is a solution to
geodesic deviation equation, for any geodesic.
These features of spacetimes enable us to discuss particle motion analytically.
It is therefore natural that many authors have attempted to generalize the notion of
spacetime symmetry in a large variety of ways.

One attempt to generalization of spacetime symmetry is to generalize 
the vector fields described above to higher-rank tensor fields.
Symmetric and antisymmetric tensor generalizations of (C)KVs are known as
(conformal) Killing-St\"ackel tensor fields\cite{Stackel:1895} ((C)KSTs) 
and (conformal) Killing-Yano tensor fields\cite{Bochner:1948,Yano:1952,Yano:1953,
Tachibana:1969,Kashiwada:1968,Kora:1980} ((C)KYTs) , respectively.
(C)KVs are closely related to conserved quantities along (null) geodesics, 
which are in particular polynomials of linear order in momenta. 
(C)KSTs are defined as symmetric tensor fields related to conserved quantities
along (null) geodesics which are polynomials of higher order in momenta. In contrast, KYTs are related to
antisymmetric tensor fields parallelly transported along geodesics, whose components are given by
polynomials of linear order in momenta. The norms of such parallelly transported antisymmetric tensor fields 
give rise to conserved quantities of second order in momenta. This means that the
squares of (C)KYTs construct rank-$2$ (C)KSTs, although not all rank-$2$ (C)KSTs can be decomposed into two (C)KYTs.
To be precise, a rank-$p$ CKST $K_{\mu_1\cdots \mu_p}$ is a rank-$p$ symmetric tensor field
$K_{\mu_1\cdots \mu_p}=K_{(\mu_1\cdots\mu_p)}$ satisfying the equation
\begin{align*}
 \nabla_{(\mu}K_{\nu_1\cdots \nu_p)} = g_{(\mu\nu_1}\Phi_{\nu_2\cdots\nu_p)} \,,
\end{align*}
where $\Phi_{\mu_1\cdots\mu_{p-1}}$ is a rank-($p-1$) symmetric tensor field.
In particular, $K_{\mu_1\cdots \mu_p}$ is called a KST if $\Phi_{\mu_1\cdots\mu_{p-1}}$ is vanishing.
A rank-$p$ CKYT $f_{\mu_1\cdots \mu_p}$ is a rank-$p$ antisymmetric tensor field
$f_{\mu_1\cdots \mu_p}=f_{[\mu_1\cdots\mu_p]}$ satisfying the equation
\begin{align*}
 \nabla_{(\mu}f_{\nu_1)\nu_2\cdots \nu_p} = p g_{\mu[\nu_1}\psi_{\nu_2\cdots\nu_p]} \,,
\end{align*}
where $\psi_{\mu_1\cdots\mu_{p-1}}$ is a rank-($p-1$) antisymmetric tensor field.
In particular, $f_{\mu_1\cdots \mu_p}$ is called a KYT if $\psi_{\mu_1\cdots\mu_{p-1}}$ is vanishing.

Since the Lie derivative of linear connections $\Gamma^\mu{}_{\nu\rho}$ was introduced,
affine collineation has been discussed as spacetime symmetry described by AVs which satisfy the equation
\begin{align*}
 {\cal L}_X \Gamma^\mu{}_{\nu\rho}
 = \nabla_\nu\nabla_\rho X^\mu - R_{\nu\rho}{}^\mu{}_\sigma X^\sigma = 0\,,
\end{align*}
where $R^\mu{}_{\nu\rho\sigma}$ is the Riemann curvature defined by
$(\nabla_\mu\nabla_\nu-\nabla_\nu\nabla_\mu)X^\rho=R_{\mu\nu}{}^\rho{}_\sigma X^\sigma$.
It is evident that an AV is a Jacobi field for any geodesic since
it satisfies the geodesic deviation equation
$V^\nu \nabla_\nu (V^\rho\nabla_\rho X^\mu)
= R_{\nu\rho}{}^\mu{}_\sigma V^\nu V^\rho X^\sigma$
for any tangent vector field $V^\mu$ to geodesics (i.e. $V^\mu\nabla_\mu V^\nu=0$) with affine parametrization.
Hence, AVs preserve any geodesic to another geodesic together with affine parametrization.
Alternatively, AVs are defined by the equation
\begin{align*}
\nabla_\mu \nabla_{(\nu}X_{\rho)}=0 \,. 
\end{align*}
The form of this equation, which is obtained by taking the covariant derivative of Killing equation,
is highly suggestive to generalize AVs to higher-rank tensor fields.
In Sec.\ II, we formally define tensor generalizations of AVs,
which will be called symmetric and antisymmetric affine tensor fields (abbreviated by SATs and AATs),
by replacing the part of Killing equation with KST and KYT equations.
Then, we investigate their properties about geodesics.

It is also worth discussing what are conditions for the existence of AATs,
what kinds of spacetimes admit AATs and, if AATs exist, how many AATs can exist.
To discuss these issues, we make a use of integrability conditions for AATs,
which are actually discussed in terms of parallel sections of a certain vector bundle.
The method was found in order to show that KVs, HVs, CKVs and AVs form vector spaces of finite dimensions and
their dimensions are bounded by $n(n+1)/2$, $n(n+1)/2 + 1$,
$(n+1)(n+2)/2$ and $n(n+1)$, respectively, and has been used in the study of
(C)KSTs\cite{Thomas:2006,Michel:2014} and (C)KYTs\cite{Semmelmann:2002,David:2004}.
Recently, with integrability conditions,
KYTs were elaborated on various spacetimes in four and five dimensions\cite{Houri:2014b}.

This paper is organized as follows.
In Sec.\ \ref{sec:affine_tensors}, tensor generalizations of affine vector fields 
called SATs and AATs are introduced.
After giving a brief review the properties of SATs,
which have been studied in earlier works\cite{Caviglia:1982,Cook:2009}, 
we investigate the properties of AATs to explore an application to physics.
Then, we show that AATs are related to parallelly transported
antisymmetric tensor fields along geodesics,
which also implies that there are conserved quantities along geodesics.
In Sec.\ \ref{sec:integrability}, we discuss a possibility of the existence of AATs.
First, we show that AATs are one-to-one corresponding to parallel sections of a certain vector bundle.
Then, we compute the integrability conditions for the parallel sections
and provide the upper bound on the maximum number of linear independent AATs.
In Sec.\ \ref{sec:example}, some examples of spacetimes admitting AATs are provided.
Sec.\ \ref{sec:conclusion} is devoted to conclusion.

\section{Affine tensor fields}
\label{sec:affine_tensors}

\subsection{Symmetric affine tensor fields}
The symmetric affine tensor fields (SATs) have already been discussed
in previous works\cite{Caviglia:1982,Cook:2009}.
In this section, we review the definition of SATs and their properties.
Throughout, we will adopt this name of SATs to distinguish symmetric and antisymmetric ones,
although the authors of previous works\cite{Caviglia:1982,Cook:2009} simply called them affine tensor fields.

\begin{definition}[Symmetric affine tensor fields\cite{Caviglia:1982,Cook:2009}]\label{def:SAT}
A symmetric tensor field $K_{\mu_1\cdots \mu_p}$ of rank $p$
is called a rank-$p$ symmetric affine tensor field if it satisfies the equation
\begin{align}
	\nabla_\mu \nabla_{(\nu}K_{\rho_1\cdots \rho_p)} = 0 \,. \label{eq:SAT}
\end{align}
\end{definition}

The significance of SATs is that they are one-to-one corresponding to Jacobi fields
which are written in the form
\begin{align}
	X^\mu = K^\mu{}_{\nu_1\nu_2\cdots \nu_{p-1}}
	V^{\nu_1}V^{\nu_2}\cdots V^{\nu_{p-1}} \,, \label{eq:JacobiField}
\end{align}
where $V^\mu$ is a tangent vector field of geodesics with affine parametrization.
If one constructs a vector field $X^\mu$ from an SAT $K_{\mu_1\cdots \mu_p}$ by Eq.\ \eqref{eq:JacobiField},
$X^\mu$ satisfies the geodesic deviation equation
\begin{align}
V^\nu \nabla_\nu (V^\rho\nabla_\rho X^\mu)
= R_{\nu\rho}{}^\mu{}_\sigma V^\nu V^\rho X^\sigma \,. \label{eq:GDE}
\end{align}
Conversely, if $X^\mu$ given by Eq.\ \eqref{eq:JacobiField} is a solution to Eq.\ \eqref{eq:GDE},
$K_{\mu_1\cdots \mu_p}$ is an SAT.

The defining equation \eqref{eq:SAT} can be rewritten as
\begin{align}
	\nabla_{(\mu}K_{\nu_1\cdots \nu_p)} = L_{\mu\nu_1\cdots\nu_p} \,, \label{eq:SAT2}
\end{align}
where $L_{\mu_1\cdots\mu_{p+1}}$ is a covariantly constant rank-($p+1$) symmetric tensor field, i.e. 
$\nabla_\mu L_{\nu_1\cdots \nu_{p+1}} = 0$ and
$L_{\mu_1\cdots \mu_{p+1}} = L_{(\mu_1\cdots \mu_{p+1})}$.
With this equation, symmetric homothetic tensor fields (SHTs) are defined as SATs
such that $L_{\mu_1\cdots\mu_{p+1}}$ is proportional to the metric $g_{\mu\nu}$.

\begin{definition}[Symmetric homothetic tensor fields\cite{Cook:2009}]\label{def:SHT}
A symmetric tensor field $K_{\mu_1\cdots \mu_p}$ of rank $p$
is called a rank-$p$ symmetric homothetic tensor field if it satisfies the equation
\begin{align*}
	\nabla_{(\mu}K_{\nu_1\cdots \nu_p)}
	= g_{(\mu\nu_1}\Phi_{\nu_2\cdots\nu_p)} \,, 
\end{align*}
where $\Phi_{\mu_1\cdots\mu_{p-1}}$
is a covariantly constant rank-$(p-1)$ symmetric tensor field.
\end{definition}


Given a (C)KST of rank $p$, denoted by $K_{\mu_1\cdots \mu_p}$,
one can construct a conserved quantity $Q=K_{\mu_1\cdots \mu_p}V^{\mu_1}\cdots V^{\mu_p}$
along (null) geodesics with tangent $V^\mu$: $V^\mu\nabla_\mu Q=0$.
In contrast, given an SAT of rank $p$, $K_{\mu_1\cdots \mu_p}$ again,
$Q$ is not in general conserved along a geodesic, but satisfies the equation
\begin{align}
	V^\mu \nabla_\mu (V^\nu \nabla_\nu Q) = 0 \,. \label{eq:KeyEq}
\end{align}
In Hamiltonian formalism, since we have $V^\mu\nabla_\mu F = \{H,F\}$ for a function $F$,
where $\{~,~\}$ is the Poisson bracket and
$H$ is the Hamiltonian for geodesics,
$H=(1/2)g_{\mu\nu}V^\mu V^\nu$,
the above equation \eqref{eq:KeyEq} is equivalent to $\{H,C\}=0$ with $C = \{H,Q\}$.
We thus obtain a conserved quantity $C$ along geodesics,
which is given by
\begin{align}
	C = L_{\mu_1\cdots \mu_{p+1}}V^{\mu_1}\cdots V^{\mu_{p+1}} \,, \label{eq:COM}
\end{align}
where $L_{\mu_1\cdots \mu_{p+1}}$ was given in Eq.\ \eqref{eq:SAT2}.

\begin{prop}[Caviglia, Zordan and Salmistaro\cite{Caviglia:1982}]\label{prop:SAT1}
Let $K_{\mu_1\cdots \mu_p}$ be a rank-$p$ SAT.
Then, $L_{\mu_1\cdots \mu_{p+1}}=\nabla_{(\mu_1}K_{\mu_2\cdots \mu_{p+1})}$ is a
covariantly constant KST of rank $p+1$ and, hence, $C$ given by Eq.\ \eqref{eq:COM} is a conserved quantity
along geodesics with tangent $V^\mu$.
\end{prop}

Recall that for two (C)KVs $\xi^\mu$ and $\eta^\mu$,
their symmetric tensor product $K_{\mu\nu}=\xi_{(\mu}\eta_{\nu)}$ is a rank-$2$ (C)KST.
Generally, the symmetric tensor products of two (C)KSTs are (C)KSTs.
Now, we fail to construct an SAT from two proper SATs in the similar way.
However, the following property is obtained.

\begin{prop}[Cook and Dray\cite{Cook:2009}]\label{prop:SAT2}
Let $\xi^\mu$ be an AV and $\eta^\mu$ be a covariantly constant vector field.
Then, the symmetric product $K_{\mu\nu}=\xi_{(\mu}\eta_{\nu)}$ is a rank-2 SAT.
Generally, the symmetric tensor product,
\begin{align*}
	K_{\mu_1\cdots \mu_{p+q}}
	= \xi_{(\mu_1\cdots \mu_p}\eta_{\mu_{p+1}\cdots \mu_{p+q})} \,, 
\end{align*}
of an SAT of rank $p$, $\xi_{\mu_1\cdots \mu_p}$,
and a covariantly constant tensor field of rank $q$, $\eta_{\mu_1\cdots \mu_q}$, is an SAT of rank $p+q$.
\end{prop}

An SAT is said to be reducible if it is decomposed into the symmetric tensor product of two SATs.
Given a reducible SAT, the corresponding conserved quantity along geodesics, given by Eq.\ \eqref{eq:COM}, is also reducible.

\subsection{Antisymmetric tensor generalization}
\label{subsec:AAT}
We define antisymmetric affine tensor fields (AATs) as follows:

\begin{definition}[Antisymmetric affine tensor fields]\label{def:AAT}
An antisymmetric tensor field $f_{\mu_1\cdots \mu_p}$ of rank $p$
is called a rank-$p$ antisymmetric affine tensor field if it satisfies the equation
\begin{align}
	\nabla_\mu \nabla_{(\nu}f_{\rho_1)\rho_2\cdots \rho_p} = 0 \,. \label{eq:AAT}
\end{align}
In particular, $f_{\mu_1\cdots \mu_p}$ is said to be closed if it is a closed form
satisfying $\nabla_{[\mu}f_{\nu_1\cdots \nu_p]}=0$.
\end{definition}

As is the case in SAT, the defining equation \eqref{eq:AAT} can be rewritten in the form
\begin{align}
	\nabla_{(\mu}f_{\nu_1)\nu_2\cdots \nu_p} = N_{\mu\nu_1\nu_2\cdots\nu_p} \,, \label{eq:AAT_2}
\end{align}
where $N_{\nu_1\cdots\nu_{p+1}}$ is a covariantly constant rank-$(p+1)$ tensor field 
satisfying the conditions
\begin{subequations}
\begin{align}
	& \nabla_\mu N_{\nu_1\cdots \nu_{p+1}} = 0 \,, \label{eq:AAT_3_1} \\
	& N_{\mu_1\cdots \mu_{p+1}} = N_{(\mu_1\mu_2)[\mu_3\cdots \mu_{p+1}]} \,, \label{eq:AAT_3} \\
        & N_{(\mu_1\mu_2\mu_3)\mu_4\cdots \mu_{p+1}} = 0 \,.
\end{align}
\end{subequations}
Notice that if we consider the case where $N_{\nu_1\cdots\nu_{p+1}}=0$,
$f_{\mu_1\cdots \mu_p}$ becomes a KYT.
Antisymmetric homothetic tensor fields (AHTs) are defined as AATs
such that $N_{\mu_1 \cdots \mu_{p+1}}$ is written by terms proportional to $g_{\mu\nu }$.

\begin{definition}[Antisymmetric homothetic tensor fields]\label{def:AHT}
An antisymmetric tensor field $f_{\mu_1\cdots \mu_p}$ of rank $p$
is called a rank-$p$ antisymmetric homothetic tensor field if it satisfies the equation
\begin{align*}
	\nabla_{(\mu}f_{\nu_1)\cdots \nu_p} = p g_{\mu[\nu_1}\psi_{\nu_2\cdots\nu_p]} \,,
\end{align*}
where $\psi_{\mu_1\cdots\mu_{p-1}}$
is a covariantly constant antisymmetric tensor field of rank $p-1$.
\end{definition}

From Eq.\ \eqref{eq:AAT_2} and the same equation with cyclic permutations of all the indices, we obtain
\begin{align}
\nabla_\mu f_{\nu_1\cdots \nu_p}
= \nabla_{[\mu}f_{\nu_1\cdots \nu_p]} + \frac{2p}{p+1} N_{\mu [\nu_1\cdots \nu_p]} \,. \label{eq:AAT_4}
\end{align}
Further taking the covariant derivative of Eq.\ \eqref{eq:AAT_4}, we obtain
\begin{align}
\nabla_\mu \nabla_{[\nu_1} f_{\nu_2 \cdots \nu_{p+1}]}
=& (p+1)R_{\mu [\nu_1\nu_2}{}^\rho f_{|\rho|\nu_3 \cdots \nu_{p+1}]} \,. \label{eq:AAT_5}
\end{align}
This equation will be used in Sec. \ref{sec:integrability} to derive the integrability conditions.

In the case that $f_{\mu_1\cdots \mu_p}$ is a closed AAT, the first term of the right-hand side of Eq.\ \eqref{eq:AAT_4} is dropped 
and it reduces to the form
\begin{align*}
	&\nabla_\mu f_{\nu_1\cdots \nu_p}
= \frac{2p}{p+1} N_{\mu [\nu_1\cdots \nu_p]} \,, 
\end{align*}
with a condition
\begin{align}
	&\nabla_\mu \nabla_\nu f_{\rho_1\cdots \rho_p} = 0 \,, \label{eq:Closed_AAT3}
\end{align}
which is derived from Eq. \eqref{eq:AAT_3_1}.

In anology with SATs, we have the following two properties of AATs:

\begin{prop}\label{prop:AAT1}
Let $f_{\mu_1\cdots \mu_p}$ be an AAT of rank $p$ and
$h_{\mu_1\cdots \mu_p}$ be a covariantly constant antisymmetric tensor field of the same rank $p$.
Then, $K_{\mu\nu}=f_{(\mu|\rho_1\cdots \rho_{p-1}|}h_{\nu)}{}^{\rho_1\cdots\rho_{p-1}}$ is a rank-$2$ SAT.
\end{prop}
{\it Proof.} For the tensor field $K_{\mu\nu}$, it is calculated that
\begin{align*}
\nabla_{\mu}K_{\nu\rho}
=& \frac{1}{2}(h_{\rho}{}^{\sigma_1\cdots\sigma_{p-1}} \nabla_\mu f_{\nu\sigma_1\cdots \sigma_{p-1}}
+ h_{\nu}{}^{\sigma_1\cdots\sigma_{p-1}} \nabla_\mu f_{\rho\sigma_1\cdots \sigma_{p-1}} ) \,,
\end{align*}
hence we have
\begin{align*}
\nabla_{(\mu}K_{\nu\rho)}
=& \frac{1}{3}(h_{\mu}{}^{\sigma_1\cdots\sigma_{p-1}} \nabla_{(\nu} f_{\rho)\sigma_1\cdots \sigma_{p-1}}
+ h_{\nu}{}^{\sigma_1\cdots\sigma_{p-1}} \nabla_{(\rho} f_{\mu)\sigma_1\cdots \sigma_{p-1}} \nonumber\\
& + h_{\rho}{}^{\sigma_1\cdots\sigma_{p-1}} \nabla_{(\mu} f_{\nu)\sigma_1\cdots \sigma_{p-1}} ) \,.
\end{align*}
Taking the covariant derivative of this equation, the right-hand side vanishes.
Then we obtain $\nabla_\sigma\nabla_{(\mu}K_{\nu\rho)}=0$,
which means that $K_{\mu\nu}$ is a rank-$2$ SAT. \hfill$\square$

\begin{prop}\label{prop:AAT2}
Let $f_{\mu_1\cdots \mu_p}$ be a closed AAT of rank $p$ and
$h_{\mu_1\cdots \mu_q}$ be a covariantly constant antisymmetric tensor field of rank $q$.
Then, the antisymmetric tensor product of the two,
\begin{align*}
k_{\mu_1\cdots\mu_{p+q}}=\frac{(p+q)!}{p!q!} f_{[\mu_1\cdots \mu_p}h_{\mu_{p+1}\cdots \mu_{p+q}]} \,,
\end{align*}
is a closed AAT of rank $p+q$.
\end{prop}
{\it Proof.} The closedness of $k_{\mu_1\cdots\mu_{p+q}}$ is evident
since it is constructed by the antisymmetric tensor product of two closed forms.
Taking the covariant derivative of $k_{\mu_1\cdots\mu_{p+q}}$ twice and 
using the conditional equation \eqref{eq:Closed_AAT3} for closed AATs,
we find $\nabla_\mu\nabla_\nu k_{\rho_1\cdots\rho_{p+q}}=0$, 
which implies that $k_{\mu_1\cdots\mu_{p+q}}$ satisfies the defining equation \eqref{eq:AAT} of AATs.
Hence, $k_{\mu_1\cdots\mu_{p+q}}$ is a closed AAT.
\hfill$\square$\\


Suppose that $f_{\mu\nu}$ is a rank-$2$ AAT and $V^\mu$
is tangent to geodesics with affine parametrization, $V^\mu\nabla_\mu V^\nu=0$.
Then, we obtain a vector field $X^\mu=V^\nu f_\nu{}^\mu$
which is orthogonal to $V^\mu$, $X_\mu V^\mu=0$, and which satisfies the equation
\begin{align*}
V^\nu\nabla_\nu X^\mu = N_{\nu\rho}{}^\mu V^\nu V^\rho \,. 
\end{align*}
Hence, we find that if $N_{\mu\nu}{}^\rho V^\mu V^\nu$ is vanishing for a geodesic,
$X^\mu$ is parallelly transported along the geodesic.
If and only if $N_{\mu\nu}{}^\rho$ is vanishing (namely, $f_{\mu\nu}$ is a KYT),
it is possible to construct a parallelly ransported vector field $X^\mu$ for each geodesic.
If $f_{\mu\nu}$ is not a KYT but a proper AAT,
$X^\mu$ satisfies the equation
\begin{align*}
 V^\rho\nabla_\rho (V^\nu\nabla_\nu X^\mu) = 0 \,, 
\end{align*}
which means that the vector field
\begin{align*}
 Y^\mu= V^\nu\nabla_\nu X^\mu = V^\nu V^\rho \nabla_{(\nu}f_{\rho)}{}^\mu \:, 
\end{align*}
is parallelly transported along a geodesic with tangent $V^\mu$.
In the similar fashion, we obtain the following proposition for AATs of any rank.

\begin{prop}\label{prop:AAT4}
Let $f_{\mu_1\cdots\mu_p}$ be an AAT of rank $p$.
Then, the antisymmetric tensor field $Y^{\mu_1\cdots \mu_{p-1}}$ given by
\begin{align}
Y^{\mu_1\cdots\mu_{p-1}} = V^\nu V^\rho \nabla_{(\nu}f_{\rho)}{}^{\mu_1\cdots\mu_{p-1}} \:, \label{eq:Def_Y}
\end{align}
is orthogonal to $V^\mu$, $V^\nu Y_\nu{}^{\mu_1\cdots\mu_{p-2}}= 0$, and it is parallelly
transported along geodesics with tangent $V^\mu$,
\begin{align*}
 V^\nu\nabla_\nu Y^{\mu_1\cdots \mu_{p-1}} = 0 \,.
\end{align*}
\end{prop}

Given a tensor field $Y^{\mu_1\cdots \mu_{p-1}}$ which is parallelly transported along geodesics,
the norm of $Y^{\mu_1\cdots \mu_{p-1}}$ gives a conserved quantity along the geodesics.
\begin{prop}
Let $f_{\mu_1\cdots\mu_p}$ be an AAT of rank $p$. Then,
the norm of the tensor field $Y^{\mu_1\cdots \mu_{p-1}}$ given by Eq.\ \eqref{eq:Def_Y},
\begin{align}
C =& Y_{\mu_1\cdots\mu_{p-1}}Y^{\mu_1\cdots\mu_{p-1}} \nonumber\\
=& \nabla_{(\nu}f_{\rho)\mu_1\cdots\mu_{p-1}}
\nabla_{(\sigma}f_{\kappa)}{}^{\mu_1\cdots\mu_{p-1}} V^\nu V^\rho V^\sigma V^\kappa \,,
\label{eq:4th_conserved}
\end{align}
is a conserved quantity along geodesics with tangent $V^\mu$.
\end{prop}

The conserved quantity obtained above is a polynomial of forth order in $V^\mu$,
which implies that $K_{\mu\nu\rho\sigma} = \nabla_{(\mu}f_{\nu|\kappa_1\cdots\kappa_{p-1}|}
\nabla_{\rho}f_{\sigma)}{}^{\kappa_1\cdots\kappa_{p-1}}$ is a covariantly constant rank-$4$ KST.

\section{Integrability conditions for antisymmetric affine tensor fields}
\label{sec:integrability}
In the previous section we have seen that if AATs exist on a spacetime,
there are some remarkable properties.
However, whether AATs exist or not on a given spacetime is another problem.
To answer this question, it is useful to consider the integrability conditions, 
for AATs to exist. 
The integrability conditions for AATs are obtained
by promoting smoothness of AATs and as a result give
strong constraints for the curvature of spacetimes.

We first show that Eq.\ \eqref{eq:AAT}
is equivalent to the following system of PDEs for $f_{\mu_1 \cdots \mu_p}$,
$F_{\mu_1 \cdots \mu_{p+1}}$ and $N_{\mu_1 \cdots \mu_{p+1}}$,
\begin{subequations}
\begin{align}
& \nabla_\mu f_{\nu_1 \cdots \nu_p}
= F_{\mu \nu_1 \cdots \nu_p} + \frac{2p}{p+1} N_{\mu [\nu_1 \cdots \nu_p]}\,, \label{eq:curvature_condition1-1}\\
& \nabla_\mu F_{\nu_1 \cdots \nu_{p+1}} 
= (p+1) R_{\mu [\nu_1\nu_2}{}^\rho f_{|\rho|\nu_3 \cdots \nu_{p+1}]}\,, \label{eq:curvature_condition1-2}\\
& \nabla_\mu N_{\nu_1 \cdots \nu_{p+1}} = 0\,, \label{eq:curvature_condition1-3}
\end{align}
\end{subequations}
where we are assuming that $f_{\mu_1 \cdots \mu_p}=f_{[\mu_1 \cdots \mu_p]}$,
$F_{\mu_1 \cdots \mu_{p+1}} = F_{[\mu_1 \cdots \mu_{p+1}]}$,
$N_{\mu_1\cdots \mu_{p+1}}=N_{(\mu_1\mu_2)[\mu_3\cdots \mu_{p+1}]}$
and $N_{(\mu_1\mu_2\mu_3)\mu_4\cdots \mu_{p+1}} = 0$.
Since it is clear from Eqs.\ \eqref{eq:AAT_2}--\eqref{eq:AAT_5} that
Eqs.\ \eqref{eq:curvature_condition1-1}--\eqref{eq:curvature_condition1-3} hold
for a rank-$p$ AAT $f_{\mu_1 \cdots \mu_p}$, we show the inverse here.
Starting from Eq.\ \eqref{eq:curvature_condition1-1}, we see that
\begin{align}
\nabla_{(\mu}f_{\nu_1)\nu_2\cdots \nu_p}
=& \frac{p}{p+1}\left(N_{\mu [\nu_1 \cdots \nu_p]}+N_{\nu_1 [\mu \nu_2 \cdots \nu_p]}\right) \nonumber\\
=& N_{\mu\nu_1\cdots\nu_p} \,, \label{eq:AAT_7}
\end{align}
hence, with Eq.\ \eqref{eq:curvature_condition1-3}, we obtain Eq.\ \eqref{eq:AAT}.
From Eqs.\ \eqref{eq:curvature_condition1-1}, \eqref{eq:curvature_condition1-2} and \eqref{eq:AAT_7},
we further obtain $F_{\mu \nu_1\cdots\nu_{p}}= \nabla_{[\mu} f_{\nu_1 \cdots \nu_p]}$.

Now, we can regard Eqs.\ \eqref{eq:curvature_condition1-1}--\eqref{eq:curvature_condition1-3}
as the parallel equation on the vector bundle $E^p=\Lambda^pT^*M\oplus \Lambda^{p+1}T^*M\oplus \Lambda^{p,1}T^*M$.
Note that sections $f_{\mu_1 \cdots \mu_p}$, $F_{\mu_1 \cdots \mu_{p+1}}$ and $N_{\mu_1 \cdots \mu_{p+1}}$
of each vector bundle $\Lambda^pT^*M$, $\Lambda^{p+1}T^*M$ and $\Lambda^{p,1}T^*M$
satisfy the conditions $f_{\mu_1 \cdots \mu_p}=f_{[\mu_1 \cdots \mu_p]}$,
$F_{\mu_1 \cdots \mu_{p+1}} = F_{[\mu_1 \cdots \mu_{p+1}]}$,
$N_{\mu_1\cdots \mu_{p+1}}=N_{(\mu_1\mu_2)[\mu_3\cdots \mu_{p+1}]}$
and $N_{(\mu_1\mu_2\mu_3)\mu_4\cdots \mu_{p+1}} = 0$.
Namely, sections of $\Lambda^pT^*M$ and $\Lambda^{p+1}T^*M$ are $p$-forms and $(p+1)$-forms, respectively,
and sections of $\Lambda^{p,1}T^*M$ are rather complicated.
From this viewpoint, we find that AATs are one-to-one corresponding to parallel sections of $E^p$.
Hence, the dimension of the space of rank-$p$ AATs is bounded by the rank of $E^p$.
Denoting by $AAT^p(M)$ the space of rank-$p$ AATs on an $n$-dimensional spacetime $(M,g_{\mu\nu})$,
we obtain the inequality
\begin{align}
\textrm{dim}~AAT^p(M) \leq
\left(
\begin{array}{c}
n+1 \\
p+1
\end{array}
\right) \times (p+1) \,, \label{eq:AAT_bound}
\end{align}
where the equality holds if a spacetime is flat.
In 4 dimensions, the maximum numbers of AVs, rank-$2$ and rank-$3$ AATs are given by
20, 30 and 20, respectively.

We apply the covariant derivative $\nabla_\sigma$ to the both sides of Eq.\ \eqref{eq:curvature_condition1-1}
and eliminate $\nabla_\sigma F_{\mu\nu_1\cdots\nu_p}$ and $\nabla_\sigma N_{\mu[\nu_1\cdots\nu_p]}$
from the right-hand side by using Eqs.\ \eqref{eq:curvature_condition1-2} and \eqref{eq:curvature_condition1-3}.
Then, we obtain the integrability conditions
\begin{subequations}
\begin{align}
R_{\sigma\mu[\nu_1}{}^\rho f_{|\rho|\nu_2\cdots\nu_p]}
= \frac{p+1}{p}\Big(R_{\sigma[\mu\nu_1}{}^\rho f_{|\rho|\nu_2\cdots\nu_p]}
-R_{\mu[\sigma\nu_1}{}^\rho f_{|\rho|\nu_2\cdots\nu_p]}\Big) \,. \label{eq:integrability_condition1-1}
\end{align}
In the similar way, Eqs.\ \eqref{eq:curvature_condition1-2} and \eqref{eq:curvature_condition1-3}
give us the other integrability conditions
\begin{align}
R_{\sigma\mu[\nu_1}{}^\rho F_{|\rho|\nu_2\cdots\nu_{p+1}]}
=& \Big(\nabla_\sigma R_{\mu[\nu_1\nu_2}{}^\rho-\nabla_\mu R_{\sigma[\nu_1\nu_2}{}^\rho\Big)
f_{|\rho|\nu_3\cdots\nu_{p+1}]} \label{eq:integrability_condition1-2} \\
& + R_{\sigma[\nu_1\nu_2}{}^\rho F_{|\rho\mu|\nu_3\cdots\nu_{p+1}]}
- R_{\mu[\nu_1\nu_2}{}^\rho F_{|\rho\sigma|\nu_3\cdots\nu_{p+1}]} \nonumber\\
& - \frac{2p}{p+1}\Big(R_{\sigma[\nu_1\nu_2}{}^\rho N_{|\rho\mu|\nu_3\cdots\nu_{p+1}]}
- R_{\mu[\nu_1\nu_2}{}^\rho N_{|\rho\sigma|\nu_3\cdots\nu_{p+1}]}\Big) \,, \nonumber
\end{align}
and
\begin{align}
2R_{\sigma\mu(\nu_1}{}^\rho N_{|\rho|\nu_2)\nu_3\cdots\nu_{p+1}}
+(p-1)R_{\sigma\mu[\nu_3}{}^\rho N_{|\nu_1\nu_2\rho|\nu_4\cdots\nu_{p+1}]} = 0 \,. \label{eq:integrability_condition1-3}
\end{align}
\label{eq:integrability_condition}
\end{subequations}
For AATs to exist on a spacetime, the Riemann curvature is restricted
by the integrability conditions.
From the Frobenius' theorem, if these integrability conditions identically vanish,
the general solution 
to the PDEs \eqref{eq:curvature_condition1-1}--\eqref{eq:curvature_condition1-3} exists
for any initial condition,
so that there are the maximum number of AATs.
The conditions otherwise give us algebraic equations for $f_{\mu_1 \cdots \mu_p}$,
$F_{\mu_1 \cdots \mu_{p+1}}$ and $N_{\mu_1 \cdots \mu_{p+1}}$,
which restrict the space of the solutions.
If there is no soluiton to the integrability conditions, no AAT exists on the spacetime.
It should be noted that, to look for proper AATs, 
the integrability condition \eqref{eq:integrability_condition1-3} for $N_{\mu_1 \cdots \mu_{p+1}}$ 
is the most important because the spacetime admits no proper AAT 
if Eq.\ \eqref{eq:integrability_condition1-3} leads to $N_{\mu_1 \cdots \mu_{p+1}} = 0$.

Furthermore, taking the covariant derivatives of Eqs.\ \eqref{eq:integrability_condition1-1}--\eqref{eq:integrability_condition1-3},
we obtain further conditions for $f_{\mu_1 \cdots \mu_p}$,
$F_{\mu_1 \cdots \mu_{p+1}}$ and $N_{\mu_1 \cdots \mu_{p+1}}$ and, in principle,
an infinite number of conditions are obtained by taking the covariant derivatives repeatedly.
Practically speaking, however, 
we would not have to differentiate them so many times
because 
the exact numbers of solutions could be obtained in many situations 
only by the integrability conditions, or their covariant derivatives up to first or second order.
In fact, KVs and KYTs on various spacetimes in 4 and 5 dimensions were investigated by Houri and Yasui\cite{Houri:2014b},
where the exact numbers of them were obtained by the conditions up to first order.
In the following section, we investigate AATs on some physical spacetimes
by making a use of the integrability conditions.

\section{Examples}
\label{sec:example}
We give several examples of spacetimes admitting affine tensor fields.
Throughout this section, we basically restrict our discussion on SATs and AATs
to those of rank-$2$ only, although higher-rank SATs and AATs are also of great interest.
In this section, proper AATs are simply called AATs if it does not lead to serious confusion.

First, let us consider the Euclidean space in $n$ dimensions with the flat metric
\begin{align*}
ds^2 = \sum_{i=1}^n (dx^i)^2.
\end{align*}
In the flat space,
$P_i = \partial_i$ and $P_{ij} = x_i \partial_j$
are respectively KVs and proper AVs.
Namely, there are $n(n+1)$ AVs.
In particular, $P = \left.P^i\right._i$ is an HV and
$L_{ij} = P_{[ij]}$ are KVs.
Likewise, there are also $n(n^2-1)/2$ AATs:
\begin{align*}
& dx^i\wedge dx^j\:, && x^idx^j\wedge dx^k\:.
\end{align*}
All of them are linearly independent and can be obtained by the wedge products of two AVs.
In contrast, all the SATs consist of the symmetric tensor products of two AVs.
We note that any SAT has the form of either $dx^idx^j$, $x^idx^jdx^k$
or $(x^idx^j-x^jdx^i)(x^kdx^l-x^ldx^k)$.
However, not all of them are linearly independent.
Although we find that the number of linear independent SATs in the flat space 
are given by $(n+3)(n+2)(n+1)n/12$, 
we leave out the proof of it.


Next, we treat a constant curvature spacetime, such as (Anti-) de Sitter spacetime.
Such a spacetime does not in fact admit AATs of any rank.
Indeed, in a constant curvature spacetime, the Riemann tensor can be expressed as
$R_{\mu \nu \rho \lambda} = \kappa g_{\mu [\rho} g_{\lambda] \nu}$, 
where $\kappa$ is a nonzero constant.
Substituting this expression into the integrability condition \eqref{eq:integrability_condition1-3}, 
we find
\begin{align*}
& g_{\mu_3[\mu_1} N_{\mu_2] \mu_4 \mu_5 \cdots \mu_{p+3}}
+g_{\mu_4[\mu_1} N_{\mu_2] \mu_3 \mu_5 \cdots \mu_{p+3}} \\
&+(p-1) \left( 
g_{[\mu_5 |\mu_1} N_{\mu_3 \mu_4 \mu_2|\mu_6 \cdots \mu_{p+3}]}
-g_{[\mu_5 |\mu_2} N_{\mu_3 \mu_4 \mu_1|\mu_6 \cdots \mu_{p+3}]} 
\right)
= 0\:.
\end{align*}
Contracting with respect to the pairs of indices $\mu_1$, $\mu_3$ and $\mu_2$, $\mu_4$ 
in the above equation, we can obtain 
the following results:
$\left.N^{\sigma}\right._{\sigma \mu_1 \cdots \mu_p} = \left.N^{\sigma}\right._{\mu_1 \sigma \cdots \mu_p}  =0$ and 
$N_{\mu_1 \cdots \mu_{p+1}} = 0$.


As pointed out by Cook and Dray\cite{Cook:2009}, an AV and two SATs exist in G\"odel's universe and
in Einstein's static universe.
In general, the same features can be found in any spacetime with a metric of the form
\begin{align}
 ds^2 
 = g_{\mu\nu}dx^\mu dx^\nu + dz^2 \,, \label{eq:metric3}
\end{align}
where $g_{\mu\nu}$ depend only on the coordinates $x^\mu$.
For these spacetimes, $\partial_z$ is a covariantly constant KV
and $z \, \partial_z$ is an AV.
Hence, it is obtained from Proposition \ref{prop:SAT2} that $z \, dz^2$ is an SAT.
Another SAT is given by $z \, ds^2$, which is particularly an SHT.
As shown in Proposition \ref{prop:SAT1},
we can construct conserved quantities along geodesics from SATs.
Now, $p_z^2$, $p_z^3$ and $p_zH$ are constructed from
$z \, dz$, $z \, dz^2$ and $z \, ds^2$, respectively,
and they are reducible.
Unlike SATs, we cannot state that AATs exist independently of the metric functions $g_{\mu\nu}$. 
However, since AATs may occur for particular functions of g, we investigate each spacetime individually
and find neither G\"odel's universe nor Einstein's static universe having any AAT.

As an example of spacetimes with AATs, we consider pp-wave spacetimes. The
metric is written in the form
\begin{align}
	ds^2 = H(u,x,y)du^2 + 2 du dv + dx^2 + dy^2 \,, \label{eq:ppwave}
\end{align}
where $H(u,x,y)$ is a function of $u$, $x$ and $y$.
It is known that for arbitrary $H(u,x,y)$, $\partial_v$ is a covariantly constant null KV and, 
depending on the form of $H(u,x,y)$,
the number of KVs increases to the maximum by degrees.
Moreover, algebras of (C)KVs on pp-wave spacetimes have been
checked throughly in previous works (e.g. Keane and Tupper\cite{Keane:2010}). 
Nevertheless, KSTs are less known as well as KYTs, SATs, and of course AATs.

For pp-wave spacetimes for arbitrary $H(u,x,y)$, our analysis find that $u \,\partial_v$ is a closed AV. 
The dual 1-forms is given by $u \,du$.
We also find that there exist 4 AATs, which are $u \,du\wedge dx$, $u \,du\wedge dy$,
$du \wedge dx$ and $du\wedge dy$. 
The first two AATs are proper and the latter two AATs are KYTs. 
Furthermore, we find a rank-$3$ AAT $u \, du\wedge dx\wedge dy$.

From the KYTs, we obtain parallelly transported vector fields $Y_1=p_x\partial_v-p_v\partial_x$
and $Y_2=p_y \partial_v - p_v \partial_y$,
where $V_\mu= (p_u,p_v,p_x,p_y)$ is 1-form dual to a tangent vector field $V^\mu$ of geodesics.
From proposition \ref{prop:AAT4}, we also obtain parallell-transported vector fields 
$p_v Y_1$, $p_v Y_2$ and a parallelly transported antisymmetric tensor field
$Z=p_v p_y du\wedge dx - p_v p_x du\wedge dy + p_v^2 dx\wedge dy$.
The norms of those vector and tensor fields are given by $p_v^2$ and $p_v^4$,
which are reducible conserved quantities.

\section{Conclusion}
\label{sec:conclusion}
In this paper, we have formally generalized AVs to higher-rank antisymmetric tensor fields, which
have been called AATs. 
In Sec.\ \ref{sec:affine_tensors}, we have shown that AATs are related to antisymmetric
tensor fields parallelly transported along geodesics, whose components are given by polynomials of
second order in momenta \eqref{eq:Def_Y}. We have also shown that the norms of such parallelly transported antisymmetric
tensor fields give rise to conserved quantities of fourth order in momenta \eqref{eq:4th_conserved}. 
This means that we are able to construct rank-4 KSTs from AATs, although 
not all rank-$4$ KSTs can be provided in the form.

In Sec.\ \ref{sec:integrability} we have shown that spaces spanned by AATs
are vector spaces of finite dimensions.
The upper bound on the dimension is given by Eq.\ \eqref{eq:AAT_bound}.
We have also obtained the integrability conditions
\eqref{eq:integrability_condition} for AATs.
In Sec.\ \ref{sec:example}, using the integrability conditions,
we have investigated AVs and rank-$2$ AATs on some physical spacetimes.
Then, we have found that no AV nor AAT exists in spacetimes of constant cuvature,
with the exception of flat case where there are the maximum number of AVs, SATs and AATs.
If a metric has the form Eq.\ \eqref{eq:metric3}, which involves the G\"odel universe and the Einstein static universe,
a KV, an AV and two AATs exist.
We have also found several AVs and AATs in pp-wave spacetime \eqref{eq:ppwave}.
Furthermore, we have investigated AVs and AATs on black hole spacetimes in four dimensions,
but unfortunately arrived at the conclusion that Schwarzschild and Kerr spacetimes
do not admit any AV and AAT.


\begin{acknowledgments}
This work was supported by the JSPS Grant-in-Aid for Scientific Research No.\ 26$\cdot$1237.
\end{acknowledgments}

\end{document}